\documentclass[twocolumn,showpacs,preprintnumbers,superscriptaddress,amsmath,floatfix,amssymb,prd]{revtex4}
\usepackage[colorlinks=true]{hyperref}
\usepackage{graphicx}

\newcommand{\comment}[1]{}

\newcommand{\lr}[1]{ \left( #1 \right) }

\newcommand{\vev}[1]{ \langle \, #1 \, \rangle }

\newcommand{\expa}[1]{ \exp{\left( #1 \right)} }

\newcommand{\logo}{\\ \vskip -22mm
\leftline{\includegraphics[scale=0.3,clip=false]{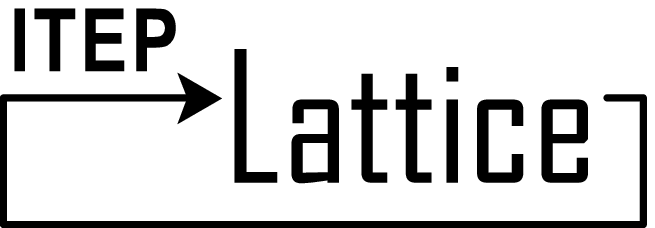}} \vskip 10mm}

\begin{document}
\sloppy
\preprint{ITEP-LAT/2008-17}

\title{Finite-temperature chiral condensate and low-lying Dirac eigenvalues in quenched $SU\lr{2}$ lattice gauge theory \logo}
\author{P. V. Buividovich}
\email{buividovich@tut.by}
\affiliation{JIPNR, National Academy of Science, 220109 Belarus, Minsk, Acad. Krasin str. 99}
\affiliation{ITEP, 117218 Russia, Moscow, B. Cheremushkinskaya str. 25}
\author{E. V. Luschevskaya}
\email{luschevskaya@itep.ru}
\affiliation{ITEP, 117218 Russia, Moscow, B. Cheremushkinskaya str. 25}
\author{M. I. Polikarpov}
\email{polykarp@itep.ru}
\affiliation{ITEP, 117218 Russia, Moscow, B. Cheremushkinskaya str. 25}
\date{September 15, 2008}
\begin{abstract}
 The spectrum of low-lying eigenvalues of overlap Dirac operator in quenched $SU\lr{2}$ lattice gauge theory with tadpole-improved Symanzik action is studied at finite temperatures in the vicinity of the confinement-deconfinement phase transition defined by the expectation value of the Polyakov line. The value of the chiral condensate obtained from the Banks-Casher relation is found to drop down rapidly at $T = T_{c}$, though not going to zero. At $T_{c}' \approx 1.5 \; T_{c} \approx 480 \, MeV$ the chiral condensate decreases rapidly one again and becomes either very small or zero. At $T < T_{c}$ the distributions of small eigenvalues are universal and are well described by chiral orthogonal ensemble of random matrices. In the temperature range above $T_{c}$ where both the chiral condensate and the expectation value of the Polyakov line are nonzero the distributions of small eigenvalues are not universal. Here the eigenvalue spectrum is better described by a phenomenological model of dilute instanton - anti-instanton gas.
\end{abstract}
\maketitle

\section{Introduction}
\label{sec:Intro}

 It is well known that in the limit of zero quark masses the classical action of QCD is invariant under the exchange of quark fields with different chirality. This classical chiral symmetry, however, appears to be broken in quantum theory. Its spontaneous breakdown gives rise to massless Goldstone particles (pions), and to nonzero expectation value of the chiral condensate $\vev{ \bar{\psi} \psi }$. Since pions are the only massless hadrons in this limit, one can describe infrared QCD using an effective chiral theory where pions are the only degrees of freedom \cite{Leutwyler:94:1,Verbaarschot:00:1}. At sufficiently high temperatures chiral symmetry is restored again, and the effective chiral theory is not valid any more, which means that at such temperatures the effective degrees of freedom are not pions. It is commonly believed that the restoration of chiral symmetry is associated with the confinement-deconfinement transition, with the chiral condensate being the order parameter. For quenched gauge theories the chiral condensate is not an exact order parameter, since in this case it does not correspond to expectation value of any physical quantity and can be defined only indirectly, via the Banks-Casher relation \cite{Banks:80:1}:
\begin{eqnarray}
\label{BanksCasher}
\Sigma \equiv |\vev{ \bar{\psi} \psi }| = \lim \limits_{\lambda \rightarrow 0} \lim \limits_{V \rightarrow \infty} \, \frac{\pi \rho\lr{\lambda}}{V}
\end{eqnarray}
where $\rho\lr{\lambda} = \vev{ \sum \limits_{i} \delta\lr{\lambda - \lambda_{i}} }$ is the density of eigenvalues $\lambda_{i}$ of the Dirac operator and $V$ is the volume of the four-dimensional box in which the theory is considered. Thus the equation (\ref{BanksCasher}) relates the chiral condensate and the density of small eigenvalues of the Dirac operator.

 In quenched $SU\lr{N}$ gauge theories a commonly used order parameter is the expectation value of the Polyakov line, which is associated with $Z_{N}$ center symmetry and which is equal to zero in the confinement phase, where center symmetry is unbroken \cite{Polyakov:78:1}. Thus the Polyakov loop and the chiral condensate are the proper order parameters for the deconfinement phase transition in the limits of infinite and zero quark masses, respectively. However, numerical simulations indicate that both the chiral condensate and the Polyakov loop can be used as approximate order parameters even when the corresponding symmetries are absent. In particular, for QCD with dynamical quarks Polyakov line changes rapidly in the vicinity of the deconfinement phase transition \cite{Karsch:94:1}. The measurements of the chiral condensate using the staggered Dirac operator in quenched $SU\lr{2}$ gauge theory has also shown that it goes to zero in the deconfinement phase \cite{Kogut:83:1}.  However, in some recent lattice studies of quenched $SU\lr{2}$ lattice gauge theory with chirally invariant lattice Dirac operator it was found that even above the deconfinement phase transition there are still some small eigenvalues $\lambda_{i}$ which yield nonzero chiral condensate in the Banks-Casher relation (\ref{BanksCasher}) \cite{Edwards:00:1, Luschevskaya:08:1}. To be more precise, it turns out that in the deconfinement phase with spontaneously broken center symmetry chiral condensate behaves differently depending on the sign of the Polyakov loop. For field configurations with positive Polyakov loop the chiral condensate eventually goes to zero at some temperature above $T_{c}$, while for configurations with negative Polyakov loops the chiral condensate stays nonzero at all temperatures which were considered \cite{Luschevskaya:08:1}. Since for $SU\lr{2}$ gauge theory with dynamical fermions at temperatures above the deconfinement phase transition positive values of Polyakov loop are favoured, at $T > T_{c}$ it seems reasonable to consider only field configurations with positive Polyakov loops in order to capture this important feature of the full QCD already in the quenched approximation \cite{Edwards:00:1, Luschevskaya:08:1}.

 The value of the chiral condensate is usually extracted from the results of lattice simulations by measuring the probability distribution $\rho\lr{\lambda}$ of the eigenvalues of the Dirac operator $\mathcal{D} = \gamma^{\mu}\lr{ \partial_{\mu} - i A_{\mu}}$ and by using the Banks-Casher relation (\ref{BanksCasher}). A much more precise method was proposed in \cite{Bitsch:98:1, Wittig:05:1}. This method is based on the fact that the properly rescaled eigenvalue distribution $\rho\lr{\lambda}$ at $\lambda \rightarrow 0$ is universal and coincides with the eigenvalue distribution of the so-called chiral ensemble of random matrices \cite{Verbaarschot:92:1, Verbaarschot:00:1}:
\begin{eqnarray}
\label{RMT_Correspondence}
\lim \limits_{V \rightarrow \infty} \frac{1}{\Sigma V} \, \rho\lr{ \frac{z}{\Sigma V}} = \rho_{S}\lr{z}
\end{eqnarray}
The form of the function $\rho_{S}\lr{z}$ depends only on the global symmetries of the Dirac operator and on the topological charge of the gauge fields. One can also consider the probability distribution $p\lr{\lambda_{min}}$ of the lowest nonzero eigenvalue of the Dirac operator, which is also universal if $\lambda_{min}$ is rescaled by the factor $\Sigma V$. Numerically it was found that for the low-lying eigenvalues the universality holds with a good precision already at not very large lattice volumes, which can be easily achieved in lattice simulations \cite{Bitsch:98:1}. In this case for $\rho\lr{\lambda}$ with $\lambda$ of order of $\lr{\Sigma V}^{-1}$ one has the following approximate equalities:
\begin{eqnarray}
\label{distrib_rmt_vs_latt}
\rho\lr{\lambda} = \frac{1}{\Sigma V}\, \rho_{S}\lr{\Sigma V \lambda}
\\
\label{distrib_lowest_rmt_vs_latt}
p\lr{\lambda_{min}} = \frac{1}{\Sigma V}\, p_{S}\lr{\Sigma V \lambda_{min}}
\end{eqnarray}
 Thus in order to find the value of $\Sigma$ one can fit the lowest edge of the numerically obtained distributions $\rho\lr{\lambda}$ or $p\lr{\lambda_{min}}$ with the universal functions $A \, \rho_{S}\lr{ c \, \lambda}$ or $A \, p_{S}\lr{c \, \lambda_{min}}$, where $A$ and $c$ are fitting parameters. The parameter $c$ is related to $\Sigma$ as $\Sigma = c \, V^{-1}$. Since such numerical procedure involves fitting lattice data with some function rather than finding a single value $\rho\lr{0}$, the precision of this method is usually significantly higher than that of the Banks-Casher relation. Finite-volume corrections are also believed to be smaller, which allows one to measure the chiral condensate using such small lattices as $4^{4}$ \cite{Bitsch:98:1, Wittig:05:1}.

\begin{table}[ht]
\centering
\begin{tabular}{|c|c|c|c|c|c|c|}
  \hline
  $N_{L}$ & $N_{T}$ & $\  \beta \  $ & $\  a, \, fm\  $ &  $\  T/T_c\  $ & \# conf. & \# eigenval. \\
  \hline
 8       & 4       & 2.93    &  0.1863     &  0.85        & 2501     & 30           \\
 8       & 4       & 2.95    &  0.1801     &  0.88        & 4977     & 30           \\
  \hline
 10      & 4       & 2.93    &  0.1863     &  0.85        & 4987     & 30           \\
 10      & 4       & 2.95    &  0.1801     &  0.88        & 4733     & 30           \\
  \hline
 12      & 6       & 3.20    &  0.1155     &  0.91        & 1908     & 20           \\
 12      & 6       & 3.23    &  0.1092     &  0.96        & 1792     & 20           \\
 12      & 6       & 3.45    &  0.0741     &  1.42        & 93       & 30           \\
  \hline
 16      & 6       & 3.20    &  0.1155     &  0.91        & 999      & 30           \\
 16      & 6       & 3.23    &  0.1092     &  0.96        & 910      & 30           \\
 16      & 6       & 3.25    &  0.1053     &  1.00        & 400      & 30           \\
 16      & 6       & 3.275   &  0.1007     &  1.04        & 519      & 30           \\
 16      & 6       & 3.325   &  0.0921     &  1.14        & 183      & 50           \\
 16      & 6       & 3.40    &  0.0807     &  1.30        & 149      & 30           \\
 16      & 6       & 3.45    &  0.0741     &  1.42        & 123      & 30           \\
 \hline
 20      & 6       & 3.16    &  0.1276     &  0.82        & 99       & 30           \\
 20      & 6       & 3.178   &  0.1219     &  0.86        & 95       & 30           \\
 20      & 6       & 3.20    &  0.1155     &  0.91        & 98       & 50           \\
 20      & 6       & 3.23    &  0.1092     &  0.96        & 299      & 50           \\
 20      & 6       & 3.275   &  0.1007     &  1.04        & 89       & 50           \\
 20      & 6       & 3.30    &  0.0960     &  1.09        & 40       & 50           \\
 20      & 6       & 3.325   &  0.0921     &  1.14        & 101      & 50           \\
 20      & 6       & 3.35    &  0.0880     &  1.19        & 123      & 50           \\
 20      & 6       & 3.40    &  0.0807     &  1.30        & 155      & 50           \\
 20      & 6       & 3.50    &  0.0681     &  1.54        & 99       & 50           \\
 20      & 6       & 3.64    &  0.0527     &  2.00        & 99       & 50           \\
 \hline
 24      & 6       & 3.23    &  0.1155     &  0.91        & 96       & 45           \\
 24      & 6       & 3.325   &  0.0921     &  0.96        & 115      & 45           \\
 24      & 6       & 3.50    &  0.0681     &  1.54        & 77       & 45           \\
 \hline
\end{tabular}
\caption{Lattice parameters which were used for simulations.}
\label{tab:sim_par}
\end{table}

 In this paper we report on our measurements of the spectrum of low-lying eigenvalues of lattice Dirac operator in quenched $SU\lr{2}$ lattice gauge theory with tadpole-improved Symanzik action (see, e.g., the expression (1) in \cite{Luschevskaya:08:1}) at finite temperatures in the vicinity of the confinement-deconfinement phase transition defined by the expectation value of the Polyakov line. In Section \ref{sec:Banks_Casher} we study the temperature dependence of the chiral condensate using the Banks-Casher relation \cite{Banks:80:1}. We demonstrate that the chiral condensate changes rapidly in the vicinity of the deconfinement phase transition, though not going to zero. Another interesting effect is a second rapid decrease of the condensate at $T \approx 1.5 \; T_{c} \approx 480 \, MeV$. In Section \ref{sec:RMT} we check the applicability of the chiral Random Matrix Theory \cite{Verbaarschot:00:1} to the spectrum of low-lying Dirac eigenvalues at finite temperatures. It is found that at $T < T_{c}$ the spectrum is universal and the Random Matrix theory can be applied, while above $T_{c}$ some other model should be used.

 Overlap Dirac operator for quarks in the fundamental representation is used to find the eigenvalue density $\rho\lr{\lambda}$ in (\ref{distrib_rmt_vs_latt}). Advantages of the overlap Dirac operator are explicit chiral symmetry at all lattice spacings and the existence of exact zero modes for field configurations with nonzero topological charge \cite{Neuberger:98:1}. The parameters of our lattice simulations are summarized in Table \ref{tab:sim_par}. We mostly use the same lattice configurations as in \cite{Luschevskaya:08:1}, thus our work can be considered as an extension of the analysis performed in that work. Specified numbers of lattice configurations are the total numbers of field configurations with all topological charges. Following \cite{Edwards:00:1, Luschevskaya:08:1}, at temperatures above $T_{c}$ we consider only gauge field configurations with positive expectation values of the Polyakov line. For quenched $SU\lr{2}$ gauge theory with tadpole-improved Symanzik action recent measurements give $\beta_{c} = 3.248(2)$ for $N_{T} = 6$, which corresponds to $T_{c} = 313.(3) \, MeV$ \cite{Bornyakov:07:1}.

\begin{figure*}[ht]
  \includegraphics[width=6cm, angle=-90]{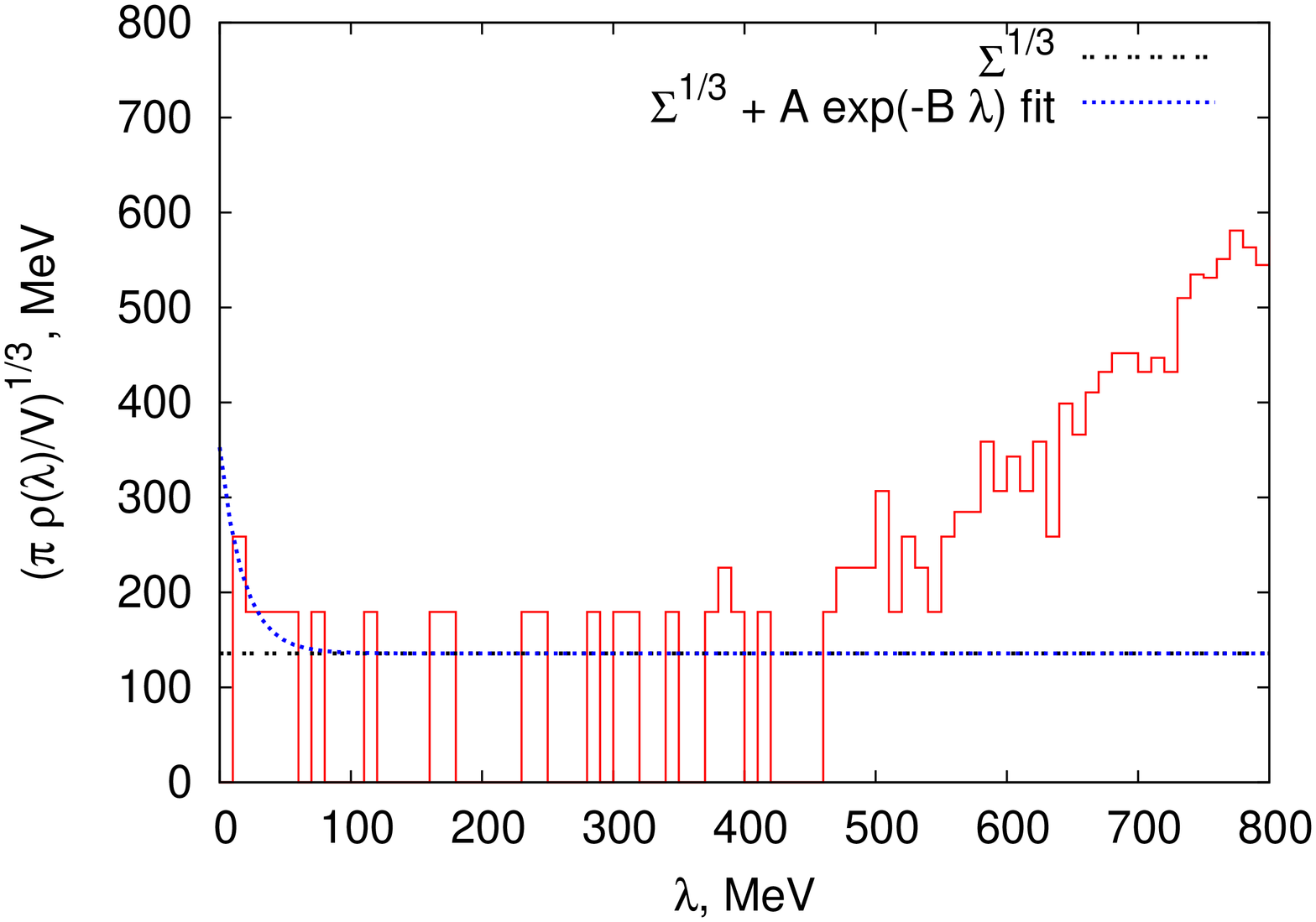} \includegraphics[width=6cm, angle=-90]{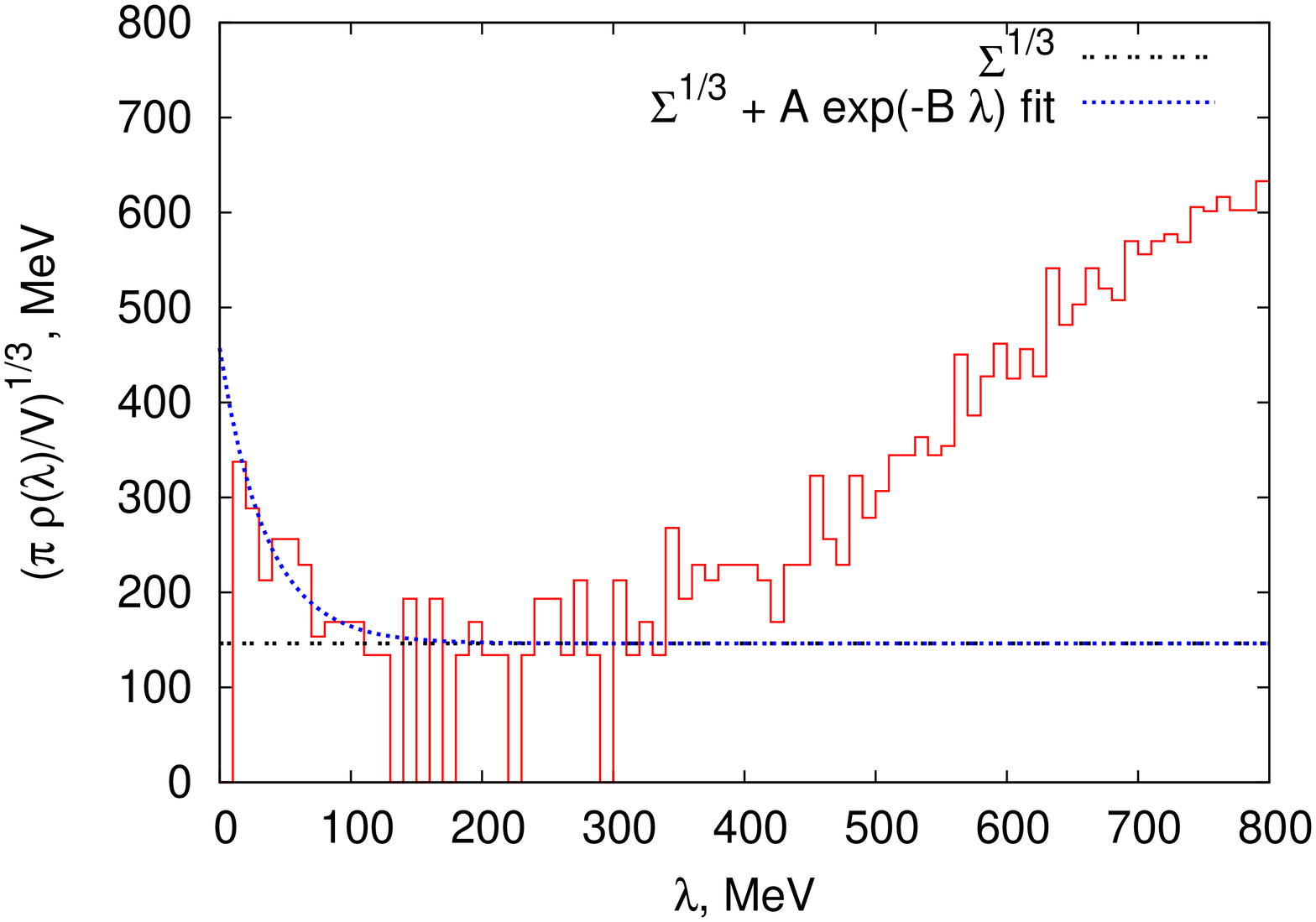}\\
  \caption{The rescaled density $\lr{\pi \rho\lr{\lambda}/V}^{1/3}$ of eigenvalues of overlap Dirac operator for $16^{3} \times 6$ lattice at $\beta = 3.40$, $T = 401 \, MeV$ (on the right) and for $20^{3} \times 6$ lattice at $\beta = 3.325$, $T = 346 \, MeV$ (on the left). Only the first $30$ and $50$ lowest eigenvalues were taken into account for the right and the left plots, respectively (see Table \ref{tab:sim_par}).}
  \label{fig:thists}
\end{figure*}

\section{Temperature dependence of the chiral condensate}
\label{sec:Banks_Casher}

 In this Section we present technical details of our measurements of the chiral condensate and discuss its temperature dependence as well as finite-volume and finite-spacing effects.

 Our results as well as the results of the previous works on the spectrum of overlap Dirac operator \cite{Edwards:00:1, Kiskis:01:1} suggest that in the vicinity of $T_{c} \approx 320 \, MeV$ the spectrum starts developing a sort of ``plateau'' for $\lambda$ in the range $100 \ldots 400 \, MeV$, which becomes wider and lower at higher temperatures. At the same time a characteristic peak gradually emerges near the origin, below approximately $100 \, MeV$. A similar feature of the spectrum of small eigenvalues in the deconfinement phase was observed in \cite{Edwards:00:1} for quenched $SU\lr{2}$ and $SU\lr{3}$ gauge theories with Wilson action and with both staggered and overlap Dirac operators and in \cite{Luschevskaya:08:1} for quenched $SU\lr{2}$ gauge theory with tadpole-improved Symanzik action and with overlap Dirac operator. The spectra of $30$ and $50$ lowest Dirac eigenvalues for $16^{3} \times 6$ lattice at $\beta = 3.40$, $T = 401 \, MeV$ and for $20^{3} \times 6$ lattice at $\beta = 3.325$, $T = 346 \, MeV$, correspondingly, are plotted on Fig. \ref{fig:thists} as an example.

\begin{figure}[ht]
  \includegraphics[width=6cm, angle=-90]{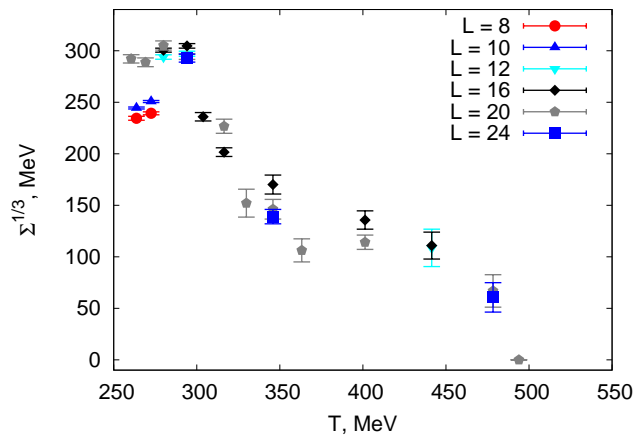}\\
  \caption{The values of the chiral condensate $\Sigma^{1/3}$ obtained from the Banks-Casher relation (\ref{BanksCasher}) at different temperatures.}
  \label{fig:condensates}
\end{figure}

\begin{figure}[ht]
  \includegraphics[width=6cm, angle=-90]{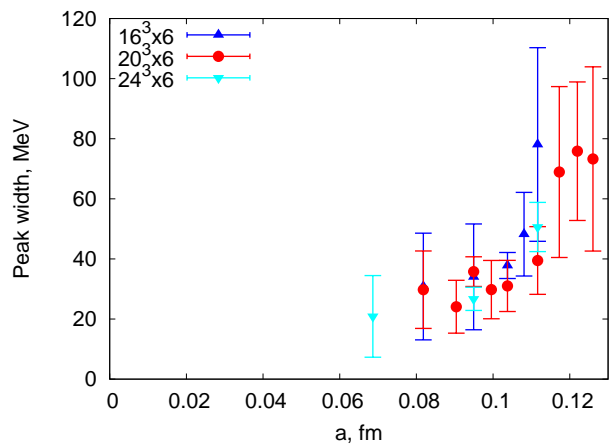}\\
  \caption{The dependence of the width of the characteristic peak in the Dirac spectra on lattice spacing.}
  \label{fig:peak_width}
\end{figure}

 For our data the height of the peak (as measured from the origin) remains practically constant at different lattice volumes and lattice spacings. Moreover, a close inspection of our results as well as of the results presented in \cite{Edwards:00:1, Kiskis:01:1} suggests that the width of this peak goes to zero in the continuum limit $a \rightarrow 0$. In order to demonstrate this, we have fitted the spectra of the Dirac operator at $\lambda < 200 \, MeV$ by the function $C + A \expa{ - B \lambda}$, where the parameter $B$ was considered as the inverse of the peak width and $C$ was the value of spectral density at the ``plateau'' (see Fig. \ref{fig:thists}). The width of the peak extracted in this way is plotted on Fig. \ref{fig:peak_width} as a function of lattice spacing for $16^{3} \times 6$, $20^{3} \times 6$ and $24^{3} \times 6$ lattices. It can be seen that the peak width rapidly drops down when the spacing is sufficiently small and the corresponding temperature is in the vicinity of the deconfinement phase transition, and approaches zero as lattice spacing becomes smaller. The widths of the peaks at different lattice sizes agree within error range. We have also analyzed the same data as the functions of four-dimensional and three-dimensional lattice volumes in physical units, but in these cases the agreement between different lattices was much worse. Thus it seems that this peak is indeed a lattice artifact, since its width goes to zero while its height remains practically constant, and in the continuum limit Dirac eigenmodes which constitute the peak turn into exact zero modes.

  Although it has been suggested in \cite{Kiskis:01:1} that the height of the peak diverges, our results in fact do not contradict those reported in \cite{Kiskis:01:1}. In \cite{Kiskis:01:1} the shape of the peak was approximated by a function of the form $\rho\lr{E} = \alpha\lr{V} \; E^{-b}$, where $E = \lambda a$ is the virtuality $\lambda$ in lattice units, $\alpha\lr{V}$ is some coefficient which increases with lattice volume $V$ and the optimal value of $b$ is $b \approx 0.80$. The dependence of the peak parameters on lattice spacing was not investigated. If one rewrites the expression for $\rho\lr{E}$ obtained in \cite{Kiskis:01:1} in physical units, one obtains $\rho\lr{\lambda} \sim \alpha\lr{V} \, a^{1 - b} \, \lambda^{-b}$. Thus for any finite value of lattice spacing the spectral density indeed diverges at $\lambda \rightarrow 0$. However, since $b < 1$, this divergence and the peak near zero disappear in the continuum limit $a \rightarrow 0$. The model of dilute instanton - anti-instanton gas used in \cite{Edwards:00:1} to describe this peak also does not involve any physical scale, which again indicates that this feature of the Dirac spectrum might be a lattice artifact.

 Taking into account the above considerations, we have assumed that at finite lattice spacing the spectral density at $\lambda \rightarrow 0$ in the Banks-Casher relation (\ref{BanksCasher}) should be replaced  by the best-fitting value, $C$, of the rescaled spectral density $\pi \rho\lr{\lambda}/V$ at the ``plateau''. These values of $C$ are shown on Fig. \ref{fig:thists} with solid horizontal lines.

 At temperatures below $400 \, MeV$ we have calculated $C$ by averaging the spectral density over $\lambda$ in the range $100 \ldots 200 \, MeV$, which is a typical extent of the ``plateau'' for such temperatures (see Fig. \ref{fig:thists}, right plot). For higher temperatures the ``plateau'' becomes significantly lower and wider. Due to limited statistics, this ``plateau'' on the histograms of $\rho\lr{\lambda}$ typically consists of only a small number of randomly distributed individual eigenvalues (see Fig. \ref{fig:thists}, left plot), and the number of eigenvalues between $100 \, MeV$ and $200 \, MeV$ strongly fluctuates. In order to reduce these fluctuations, for higher temperatures we have extended the averaging interval up to $\lambda_{max} = 400 \, MeV$. After such an extension the values of the chiral condensate above $T = 400 \, MeV$ agree very well for different lattice volumes. This is an additional argument in favor of our method to measure the chiral condensate at finite lattice spacing.

 The resulting temperature dependence of $\Sigma$ is shown on Fig. \ref{fig:condensates}. A comparison of the data for different lattice volumes shows that the finite-volume effects are significant for $8^{3}\times 4$ and $10^{3} \times 4$ lattices. Starting from $12^{3} \times 6$ lattice, for $T < T_{c} \approx 320 \; MeV$ the chiral condensate stabilizes at $\Sigma^{1/3} \approx 300 \, MeV$. At $T \approx T_{c}$ the condensate rapidly changes from $\Sigma^{1/3} \approx 300 \, MeV$ to $\Sigma^{1/3} \approx 120 \, MeV$ and then practically does not change in the temperature range $T_{c} < T < 1.5 T_{c} \approx 480 \, MeV$. This value of $\Sigma$ is also practically volume-independent for $12^{3} \times 6$, $16^{3} \times 6$, $20^{3} \times 6$ and $24^{3} \times 6$ lattices, which suggests that nonzero chiral condensate above $T_{c}$ is not a finite-volume effect. Finally, at $T \approx 1.5 \, T_{c} \approx 480 \, MeV$ $\Sigma$ quickly decreases once again. Although the chiral condensate is equal to zero for the highest temperature that we have considered ($T = 494 \, MeV$, the last point on the right on Fig. \ref{fig:condensates}), this can be just the effect of limited statistics, as discussed above. This is an interesting aspect of the theory which deserves further investigations, since in quenched theory the chiral symmetry is not a symmetry of the action and strictly speaking there are no theoretical reasons to expect that the chiral condensate should eventually go to zero at some temperature. Nevertheless, our results suggest that near $T \approx 480 \, MeV$ there is at least a sort of crossover transition associated with rapid change in the value of the chiral condensate, and that the commonly accepted deconfinement picture with zero chiral condensate at high temperatures is valid at least approximately.

\section{Comparison with chiral Random Matrix Theory and the model of dilute instanton - anti-instanton gas}
\label{sec:RMT}

 In this Section we test the universality of the spectra of low-lying Dirac eigenvalues at different temperatures by comparing them with the predictions of chiral Random Matrix Theory \cite{Verbaarschot:00:1}. Non-universal spectra are compared with the model of dilute instanton - anti-instanton gas used in \cite{Edwards:00:1}.

 The chiral ensemble of random matrices which determines the universal distributions $\rho_{S}\lr{z}$ and $p_{S}\lr{\lambda_{min}}$ for overlap Dirac operator with $SU\lr{2}$ gauge group is the chiral orthogonal ensemble, an ensemble of real random matrices with chiral structure \cite{Verbaarschot:00:1, Verbaarschot:94:2}. For such ensemble the function $\rho_{S}\lr{z}$ can be expressed in terms of some rather complicated integrals which involve Bessel functions \cite{Verbaarschot:94:2}. These integrals are rather difficult even for numerical integration, thus to obtain the function $\rho_{S}\lr{z}$ we have simply generated $5 \cdot 10^{5}$ chiral random matrices of size $50 \times 50$ for topological charges $Q = 0, 1, 2$ and used the interpolated distributions of their eigenvalues. Further increasing the size of the matrices or the number of matrices in the ensemble does not change the result within the accuracy of several tenth of percent. The functions $p_{S}\lr{z}$ are known in analytical form for topological charges $Q \neq 2$ \cite{Damgaard:01:1}:
\begin{eqnarray}
\label{lowest_density_particular}
p_{S}\lr{z} = \lr{2 + z}/4 \, \expa{ - z/2 - z^{2}/8}, \quad Q = 0
\nonumber \\
p_{S}\lr{z} = z/4 \, \expa{ - z^{2}/8}, \quad Q = 1
\end{eqnarray}
For $Q = 2$ the function $p_{S}\lr{z}$ was found numerically, similarly to $\rho_{S}\lr{z}$.

\begin{figure*}[ht]
  \includegraphics[width=6cm, angle=-90]{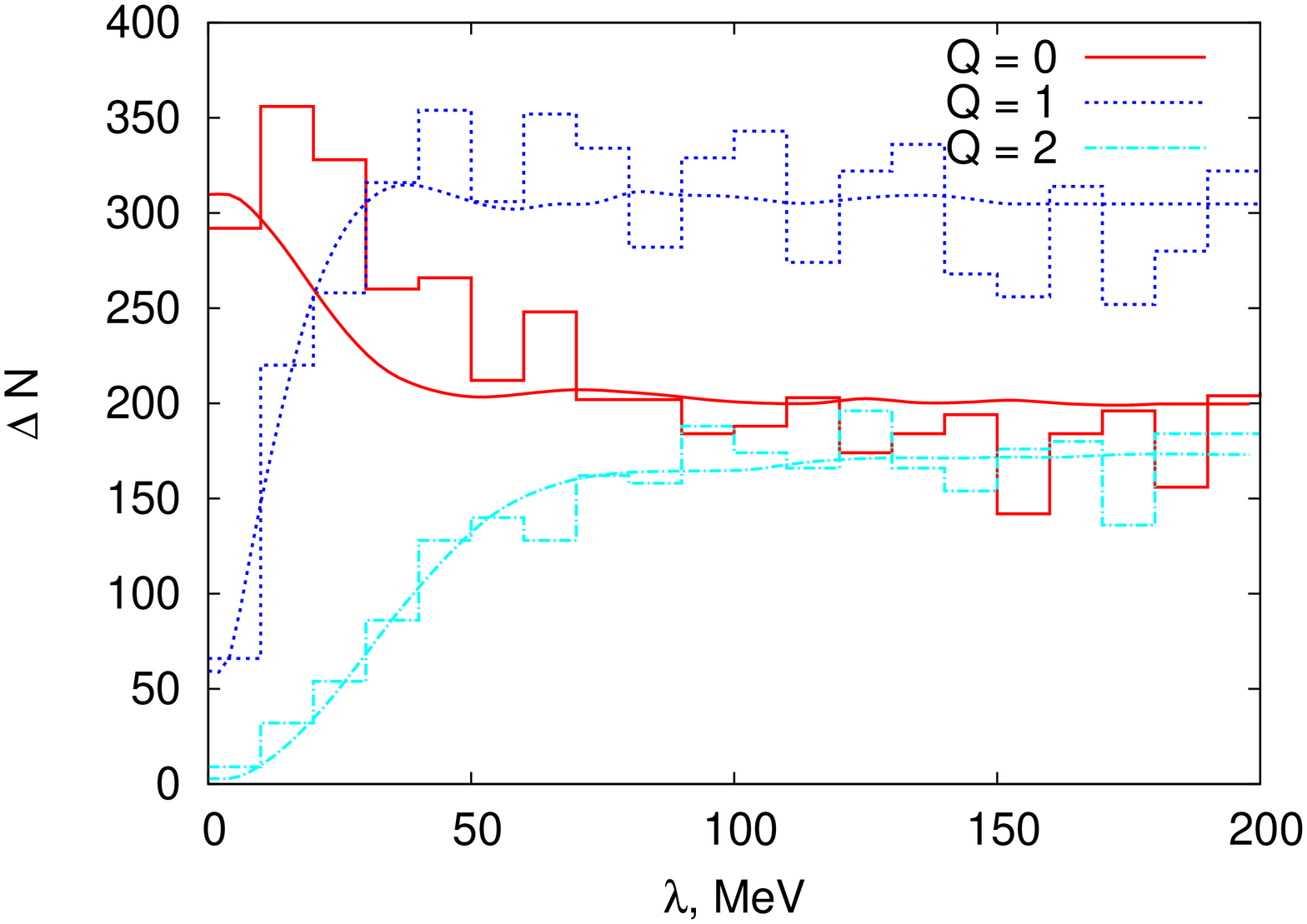}\includegraphics[width=6cm, angle=-90]{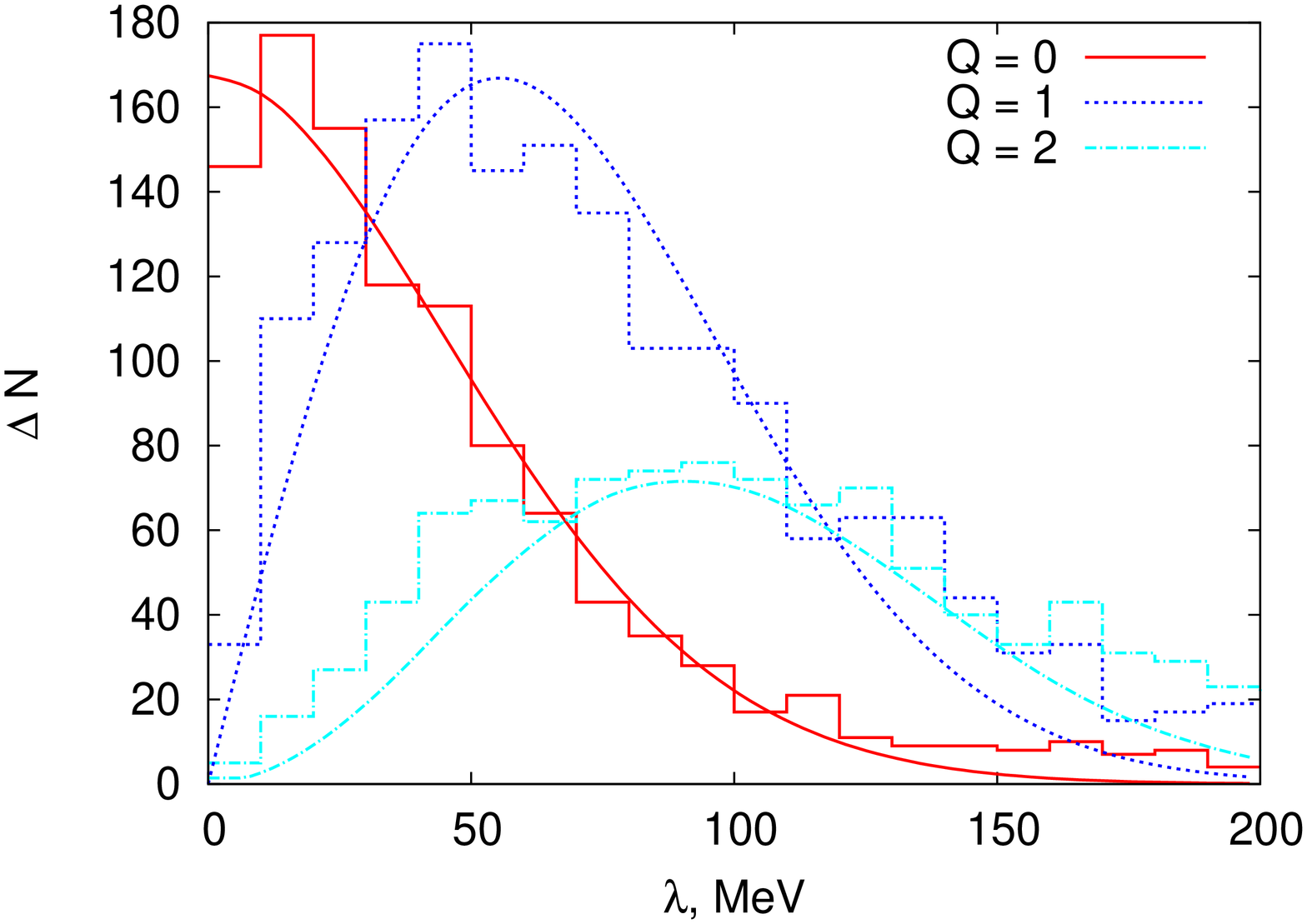}\\
  \includegraphics[width=6cm, angle=-90]{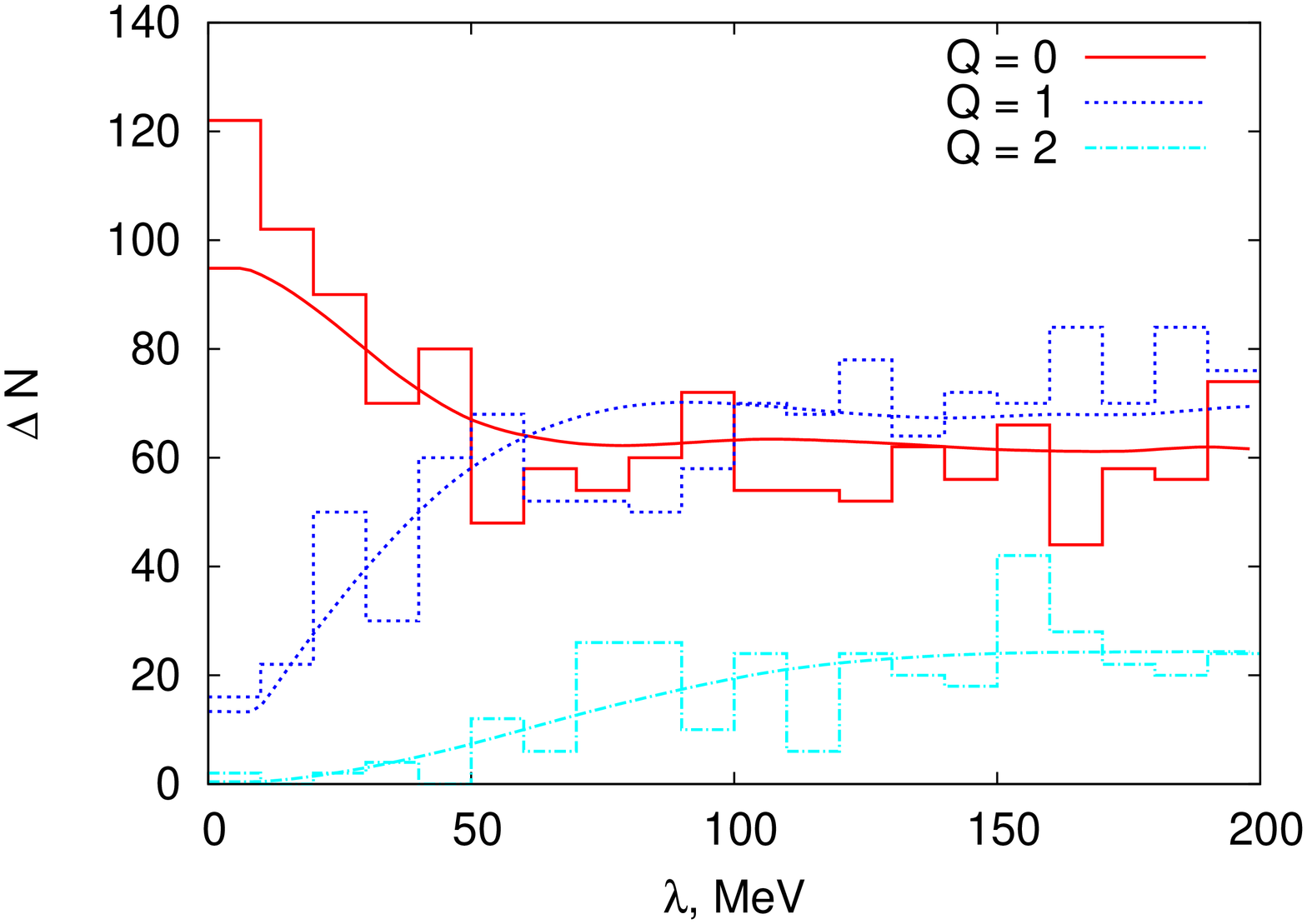}\includegraphics[width=6cm, angle=-90]{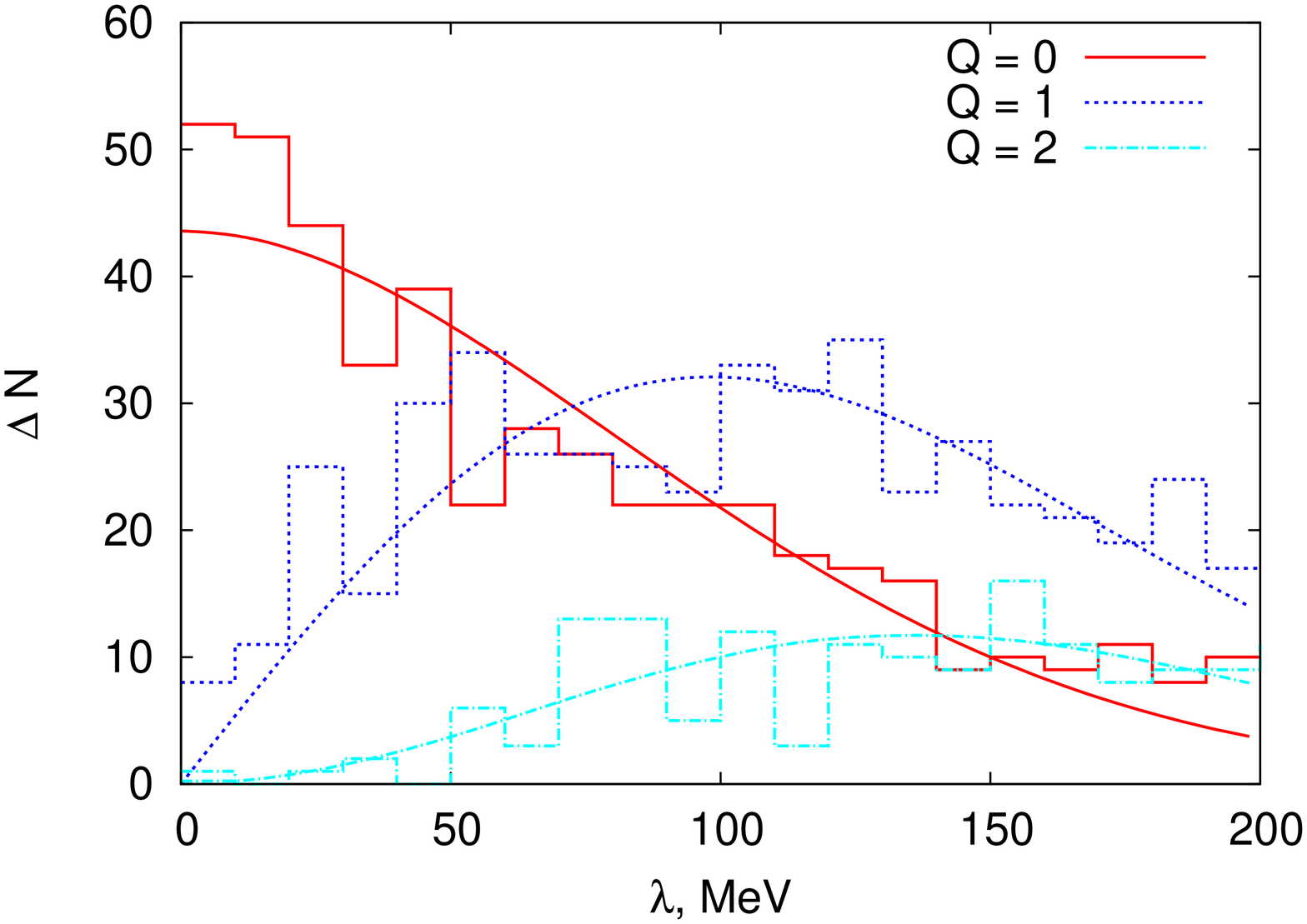}\\
  \caption{Distributions of low-lying eigenvalues of overlap Dirac operator compared with universal distributions for chiral orthogonal ensemble of random matrices. At the top: results for $10^{3} \times 4$ lattice at $\beta = 2.93$, $T = 264\, MeV$, at the bottom: results for $12^{3} \times 6$ lattice at $\beta = 3.23$, $T = 294 \, MeV$,  on the left: distributions of all eigenvalues, on the right: distributions of the lowest nonzero eigenvalues.}
  \label{fig:distribs}
\end{figure*}

\begin{table}
  \centering
  \begin{tabular}{|c|c|c|c|c|}
    \hline
    $T,\, MeV$ & $N_{L}$ & $Q$  & $\Sigma^{1/3},\, MeV$ (RMT) & $\Sigma^{1/3},\, MeV$ (B.-C.)\\
    \hline
    264        & 10      & 0 L & $221 \pm 3$                 & $244 \pm 2$  \\
    264        & 10      & 1 L & $228 \pm 3$                 &              \\
    264        & 10      & 2 L & $181 \pm 4$                 &              \\
    264        & 10      & 0 A & $207 \pm 3$                 &              \\
    264        & 10      & 1 A & $220 \pm 10$                &              \\
    264        & 10      & 2 A & $203 \pm 5$                 &              \\
    \hline
    294        & 12      & 0 L & $264 \pm 6$                 & $297 \pm 6$  \\
    294        & 12      & 1 L & $272 \pm 5$                 &              \\
    294        & 12      & 2 L & $228 \pm 6$                 &              \\
    294        & 12      & 0 A & $260 \pm 20$                &              \\
    294        & 12      & 1 A & $240 \pm 10$                &              \\
    294        & 12      & 2 A & $230 \pm 20$                &              \\
    \hline
    294        & 16      & 0 L & $308 \pm 3$                 & $304 \pm 5$  \\
    294        & 16      & 1 L & $305 \pm 6$                 &              \\
    294        & 16      & 2 L & $320 \pm 30$                &              \\
    294        & 16      & 0 A & $290 \pm 10$                &              \\
    294        & 16      & 1 A & $295 \pm 6$                 &              \\
    294        & 16      & 2 A & $300 \pm 20$                &              \\
    \hline
  \end{tabular}
  \caption{The values of the chiral condensate $\Sigma^{1/3}$ extracted by fitting the Dirac spectra obtained in lattice simulations with the universal eigenvalue distributions of chiral orthogonal ensembles with different topological charges. The values of $\Sigma^{1/3}$ obtained from Banks-Casher relation are also shown for comparison. The symbols ``L'' or ``A'' mean that the distributions of the lowest eigenvalues $p\lr{\lambda_{min}}$ or of all eigenvalues $\rho\lr{\lambda}$ were fitted.}
  \label{tab:ch_cond}
\end{table}

 As an example, the histograms of the distributions $\rho\lr{\lambda}$ and $p\lr{\lambda_{min}}$ at temperatures $T = 264 \, MeV$ ($10^{3} \times 4$ lattice, $T/T_{c} = 0.84$) and $T = 294 \, MeV$ ($12^{3} \times 6$ lattice, $T/T_{c} = 0.94$) are plotted on Fig. \ref{fig:distribs} together with the functions $\frac{1}{\Sigma V}\, \rho_{S}\lr{\Sigma V \lambda}$ and $\frac{1}{\Sigma V}\, p_{S}\lr{\Sigma V \lambda_{min}}$ at the values of $\Sigma$ which best fit the lattice data. These values are summarized in Table \ref{tab:ch_cond}, where the results obtained using the Banks-Casher relation are also included for comparison. It can be seen that for all plots in Fig. \ref{fig:distribs} the lattice data agrees with the universal distributions within the range of statistical errors. For $10^{3} \times 4$ and $12^{3} \times 6$ lattices the fits for different topological sectors give different results for $\Sigma$, and the Banks-Casher relation also yields somewhat higher value of the chiral condensate. For $16^{3} \times 6$ lattice the agreement between all measurements is much better. This indicates that for small lattices finite-volume effects are rather significant. These effects, however, influence only the value of $\Sigma$ but not the shape of the distribution of small eigenvalues, which remains universal.

\begin{figure*}[ht]
 \includegraphics[width=6cm, angle=-90]{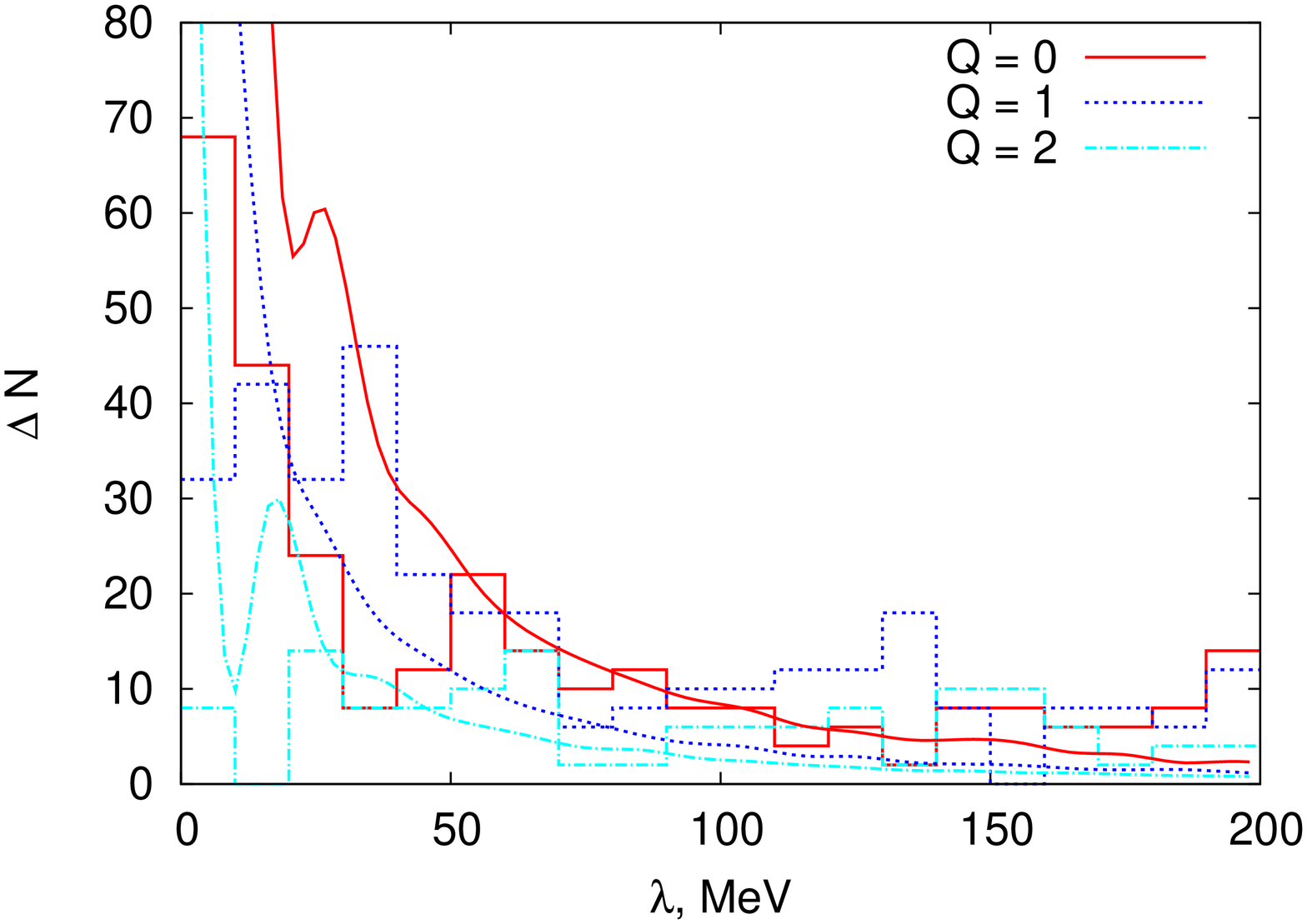}\includegraphics[width=6cm, angle=-90]{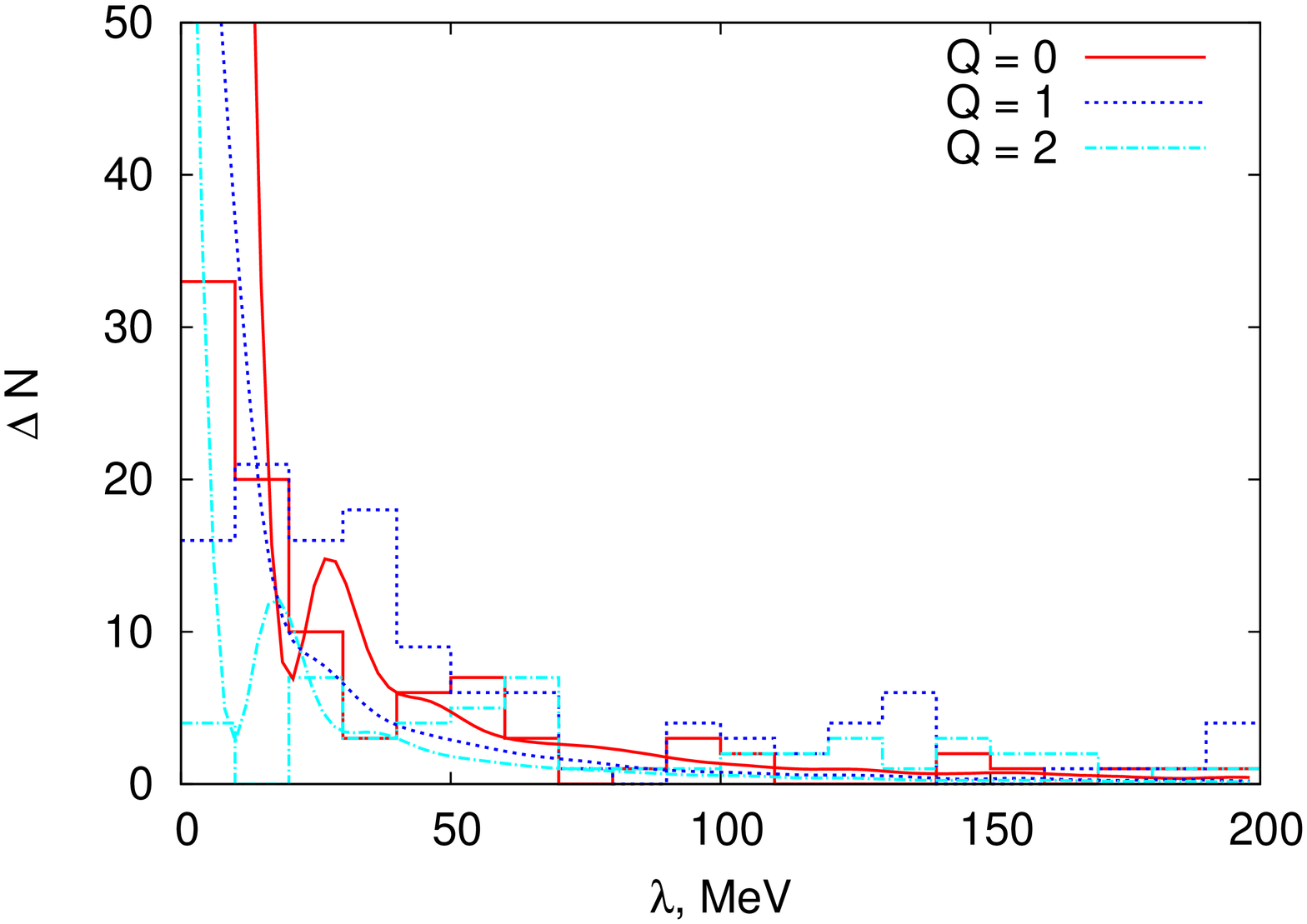}\\
 \caption{Eigenvalue distributions $\rho\lr{\lambda}$ (on the left) and $p\lr{\lambda_{min}}$ (on the right) for $16^{3} \times 6$ lattice at $\beta = 3.275$, $T = 316 \, MeV$ compared with the phenomenological model of dilute instanton - anti-instanton gas used in \cite{Edwards:00:1}.}
  \label{fig:distribs_above}
\end{figure*}

 The distributions of eigenvalues for $16^{3} \times 6$ lattice at $T = 316 \, MeV$, which is just above $T_{c}$, are plotted on Fig. \ref{fig:distribs_above}. It can be seen that these distribution are indeed not similar to those predicted by random matrix theory, and can be characterized by a strong excess of eigenvalues below approximately $100 \, MeV$. This characteristic peak in the density of the eigenvalues of the Dirac operator has already been discussed in Section \ref{sec:Banks_Casher}. Although it is most likely that this peak is a lattice artifact which disappears in the continuum limit, it seems useful to compare its shape with the one predicted by the model of dilute instanton - anti-instanton gas \cite{Edwards:00:1}. In this model instantons and anti-instantons interact at the scale of several lattice spacings rather than at a physical scale \cite{Edwards:00:1}, thus its success can serve as another argument that a strong excess of small eigenvalues at $T > T_{c}$ is a lattice artifact.

 In the model proposed in \cite{Edwards:00:1} each near-zero mode is associated either with an instanton or anti-instanton and the matrix elements of the Dirac operator between the modes $i$ and $j$ are approximated by $T_{ij} = h_{0} \expa{ - d_{ij}/D }$, where $h_{0}$ is some energy scale, $d_{ij}$ is the minimal distance in lattice units between the instanton associated with the mode $i$ and the anti-instanton associated with the mode $j$ and $D$ is a typical scale of interactions between instantons and anti-instantons induced by fermionic fields. Matrix elements between the modes associated with two instantons or two anti-instantons are assumed to be zero. With $h_{0}$ being equal to the inverse lattice spacing, the optimal value of $D$ for our data is $D = 1.3 \pm 0.1$. Eigenvalue distributions obtained from this model at the optimal value of $D$ are also plotted on Fig. \ref{fig:distribs_above}. These distributions agree well with lattice data in the range $\lambda = 20 \, MeV \, \ldots \, 100 \, MeV$, while for $\lambda \le 20 \, MeV$ the density of eigenvalues of lattice Dirac operator is significantly lower.

 Thus at high temperatures the distributions of eigenvalues within the characteristic peak near $\lambda = 0$ are well described by the model used in \cite{Edwards:00:1}, with the interaction scale between instantons being fixed in units of lattice spacing. This again suggests that the modes which constitute this peak should turn into exact zero modes in the continuum limit. This conclusion immediately rises the question about the model which could describe the spectrum of small eigenvalues of Dirac operator at not very high temperatures above $T_{c}$ in the continuum limit.

\section{Conclusions}
\label{sec:Conclusions}

 In this work we have measured the spectrum of low-lying eigenvalues of overlap Dirac operator and the value of the chiral condensate in quenched $SU\lr{2}$ lattice gauge theory with tadpole-improved Symanzik action. Our results suggest that depending on the temperature, the theory can be in the three regimes: a conventional confinement phase at $0 < T < T_{c} \approx 320 \, MeV$, for which the Polyakov line is zero and the chiral condensate is nonzero, a phase at $T > 1.5 \, T_{c} \approx 480 \, MeV$, for which the Polyakov line is nonzero and the chiral condensate is either very small or zero and which is therefore very similar to the conventional deconfinement phase, and a sort of ``transition regime'' at $T_{c} < T < 1.5 \, T_{c}$, where the Polyakov line is nonzero and the chiral condensate is still comparable with a typical hadron scale. The chiral condensate changes rapidly at the transitions between all these regimes, which suggests that even in quenched theory center and chiral symmetries are closely related \cite{Gattringer:06:1, Gattringer:08:1}, although in a somewhat more specific way than in the theory with dynamical fermions.

 The distributions of low-lying eigenvalues of the Dirac operator are universal in the confinement phase up to the critical temperature $T_{c}$, and are described by the chiral orthogonal ensemble of random matrices, as it should be \cite{Verbaarschot:00:1}. On the other hand, in the ``transition regime'' the distributions of small eigenvalues become non-universal. This could be expected, since the proof of the universality of $\rho_{S}\lr{\Sigma V \lambda }$ and $p_{S}\lr{\Sigma V \lambda_{min}}$ at small $\lambda$ is essentially based on the validity of the effective chiral theory \cite{Leutwyler:94:1, Verbaarschot:00:1}, although it is often claimed in the literature that the spectrum of small eigenvalues should be universal whenever the eigenvalue density near zero is finite \cite{Verbaarschot:00:1}. It is therefore reasonable to expect that at temperatures above $T_{c}$, at which quarks are not confined in mesons anymore and the effective chiral theory becomes invalid, these distributions become non-universal. In this temperature range the data are better described by the phenomenological model of dilute instanton - anti-instanton gas considered in \cite{Edwards:00:1}. It could be interesting to further investigate the applicability of such a model, since lattice simulations indicate that the density of topological charge is arranged into fractal-like structures which do not resemble instantons at all \cite{Horvath:03:1, Aubin:04:1, Polikarpov:05:2, Koma:05:1}. Furthermore, a typical feature of the Dirac spectrum in the ``transition regime'' which is described by the model of \cite{Edwards:00:1}, namely, the characteristic peak near $\lambda = 0$, seems to be a lattice artifact which disappears in the continuum limit. It is therefore an open question what kind of continuum model may describe the properties of small Dirac eigenvalues above $T_{c}$ and what are the objects responsible for the persistence of near-zero eigenmodes of the Dirac operator.

 Finally, at temperatures above approximately $1.5 \; T_{c}$ the density of small eigenvalues rapidly drops down once more to a value which is very close to zero. Unfortunately, due to limited statistics it is very difficult to measure this density and to establish whether it is exactly equal to zero or not. It is an intriguing open question, since for quenched theory chiral symmetry is not a symmetry of the action at all and there are no theoretical reasons to expect that the chiral condensate turns to zero at some temperature. If it really does, this would mean that in pure gauge theory there are in fact two finite-temperature phase transitions, and it is not clear what might be the origin of the second transition. It is interesting to note here that some gauge theories constructed using the gauge/gravity duality, such as the Sakai-Sugimoto model, indeed exhibit two finite-temperature phase transitions, a deconfinement one and the one associated with spontaneous breaking of chiral symmetry \cite{Aharony:07:1, Harvey:06:1}.

\begin{acknowledgments}
 The authors are grateful to M. Chernodub and E.-M. Ilgenfritz for useful discussions and to S. M. Morozov for his help with numerical implementation of the overlap operator. This work was partly supported by grants RFBR 06-02-04010-NNIO-a, RFBR 08-02-00661-a, RFBR 06-02-17012, DFG-RFBR 436 RUS, grant for scientific schools NSh-679.2008.2 and by Federal Program of the Russian Ministry of Industry, Science and Technology No 40.052.1.1.1112 and by Russian Federal Agency for Nuclear Power.
\end{acknowledgments}


\begin{thebibliography}{24}
\expandafter\ifx\csname natexlab\endcsname\relax\def\natexlab#1{#1}\fi
\expandafter\ifx\csname bibnamefont\endcsname\relax
  \def\bibnamefont#1{#1}\fi
\expandafter\ifx\csname bibfnamefont\endcsname\relax
  \def\bibfnamefont#1{#1}\fi
\expandafter\ifx\csname citenamefont\endcsname\relax
  \def\citenamefont#1{#1}\fi
\expandafter\ifx\csname url\endcsname\relax
  \def\url#1{\texttt{#1}}\fi
\expandafter\ifx\csname urlprefix\endcsname\relax\def\urlprefix{URL }\fi
\providecommand{\bibinfo}[2]{#2}
\providecommand{\eprint}[2][]{\url{#2}}

\bibitem[{\citenamefont{Leutwyler}(1994)}]{Leutwyler:94:1}
\bibinfo{author}{\bibfnamefont{H.}~\bibnamefont{Leutwyler}},
  \bibinfo{journal}{Ann. Phys.} \textbf{\bibinfo{volume}{235}},
  \bibinfo{pages}{165 } (\bibinfo{year}{1994}).

\bibitem[{\citenamefont{Verbaarschot and Wettig}(2000)}]{Verbaarschot:00:1}
\bibinfo{author}{\bibfnamefont{J.~J.~M.} \bibnamefont{Verbaarschot}}
  \bibnamefont{and} \bibinfo{author}{\bibfnamefont{T.}~\bibnamefont{Wettig}},
  \bibinfo{journal}{Ann. Rev. Nucl. Part. Sci.} \textbf{\bibinfo{volume}{50}},
  \bibinfo{pages}{343 } (\bibinfo{year}{2000}),
  \urlprefix\url{http://arxiv.org/abs/hep-ph/0003017}.

\bibitem[{\citenamefont{Banks and Casher}(1980)}]{Banks:80:1}
\bibinfo{author}{\bibfnamefont{T.}~\bibnamefont{Banks}} \bibnamefont{and}
  \bibinfo{author}{\bibfnamefont{A.}~\bibnamefont{Casher}},
  \bibinfo{journal}{Nucl. Phys. B} \textbf{\bibinfo{volume}{169}},
  \bibinfo{pages}{103 } (\bibinfo{year}{1980}).

\bibitem[{\citenamefont{Polyakov}(1978)}]{Polyakov:78:1}
\bibinfo{author}{\bibfnamefont{A.~M.} \bibnamefont{Polyakov}},
  \bibinfo{journal}{Phys. Lett. B} \textbf{\bibinfo{volume}{72}},
  \bibinfo{pages}{477} (\bibinfo{year}{1978}).

\bibitem[{\citenamefont{Karsch and Laermann}(1994)}]{Karsch:94:1}
\bibinfo{author}{\bibfnamefont{F.}~\bibnamefont{Karsch}} \bibnamefont{and}
  \bibinfo{author}{\bibfnamefont{E.}~\bibnamefont{Laermann}},
  \bibinfo{journal}{Phys. Rev. D} \textbf{\bibinfo{volume}{50}},
  \bibinfo{pages}{6954} (\bibinfo{year}{1994}),
  \urlprefix\url{http://arxiv.org/abs/hep-lat/9406008}.

\bibitem[{\citenamefont{Kogut et~al.}(1983)\citenamefont{Kogut, Matsuoka,
  Stone, Wyld, Shenker, Shigemitsu, and Sinclair}}]{Kogut:83:1}
\bibinfo{author}{\bibfnamefont{J.}~\bibnamefont{Kogut}},
  \bibinfo{author}{\bibfnamefont{H.}~\bibnamefont{Matsuoka}},
  \bibinfo{author}{\bibfnamefont{M.}~\bibnamefont{Stone}},
  \bibinfo{author}{\bibfnamefont{H.~W.} \bibnamefont{Wyld}},
  \bibinfo{author}{\bibfnamefont{S.}~\bibnamefont{Shenker}},
  \bibinfo{author}{\bibfnamefont{J.}~\bibnamefont{Shigemitsu}},
  \bibnamefont{and} \bibinfo{author}{\bibfnamefont{D.~K.}
  \bibnamefont{Sinclair}}, \bibinfo{journal}{Phys. Rev. Lett.}
  \textbf{\bibinfo{volume}{51}}, \bibinfo{pages}{869} (\bibinfo{year}{1983}),
  \urlprefix\url{http://prola.aps.org/abstract/PRL/v51/i10/p869_1}.

\bibitem[{\citenamefont{Edwards et~al.}(2000)\citenamefont{Edwards, Heller,
  Kiskis, and Narayanan}}]{Edwards:00:1}
\bibinfo{author}{\bibfnamefont{R.~G.} \bibnamefont{Edwards}},
  \bibinfo{author}{\bibfnamefont{U.~M.} \bibnamefont{Heller}},
  \bibinfo{author}{\bibfnamefont{J.}~\bibnamefont{Kiskis}}, \bibnamefont{and}
  \bibinfo{author}{\bibfnamefont{R.}~\bibnamefont{Narayanan}},
  \bibinfo{journal}{Phys. Rev. D} \textbf{\bibinfo{volume}{61}},
  \bibinfo{pages}{074504} (\bibinfo{year}{2000}),
  \urlprefix\url{http://arxiv.org/abs/hep-lat/9910041}.

\bibitem[{\citenamefont{Bornyakov et~al.}(2008)\citenamefont{Bornyakov,
  Luschevskaya, Morozov, Polikarpov, Ilgenfritz, and
  M\"{u}ller-Preussker}}]{Luschevskaya:08:1}
\bibinfo{author}{\bibfnamefont{V.~G.} \bibnamefont{Bornyakov}},
  \bibinfo{author}{\bibfnamefont{E.~V.} \bibnamefont{Luschevskaya}},
  \bibinfo{author}{\bibfnamefont{S.~M.} \bibnamefont{Morozov}},
  \bibinfo{author}{\bibfnamefont{M.~I.} \bibnamefont{Polikarpov}},
  \bibinfo{author}{\bibfnamefont{E.}~\bibnamefont{Ilgenfritz}},
  \bibnamefont{and}
  \bibinfo{author}{\bibfnamefont{M.}~\bibnamefont{M\"{u}ller-Preussker}},
  \emph{\bibinfo{title}{The topological structure of {SU(2)} gluodynamics at
  {T}$>$0 : an analysis using the {S}ymanzik action and {N}euberger overlap
  fermions}} (\bibinfo{year}{2008}),
  \urlprefix\url{http://arxiv.org/abs/0807.1980}.

\bibitem[{\citenamefont{Berbenni-Bitsch
  et~al.}(1998)\citenamefont{Berbenni-Bitsch, Jackson, Meyer, Schafer,
  Verbaarschot, and Wettig}}]{Bitsch:98:1}
\bibinfo{author}{\bibfnamefont{M.~E.} \bibnamefont{Berbenni-Bitsch}},
  \bibinfo{author}{\bibfnamefont{A.~D.} \bibnamefont{Jackson}},
  \bibinfo{author}{\bibfnamefont{S.}~\bibnamefont{Meyer}},
  \bibinfo{author}{\bibfnamefont{A.}~\bibnamefont{Schafer}},
  \bibinfo{author}{\bibfnamefont{J.~J.~M.} \bibnamefont{Verbaarschot}},
  \bibnamefont{and} \bibinfo{author}{\bibfnamefont{T.}~\bibnamefont{Wettig}},
  \bibinfo{journal}{Nucl. Phys. B - Proc. Suppl.}
  \textbf{\bibinfo{volume}{63}}, \bibinfo{pages}{820} (\bibinfo{year}{1998}),
  \urlprefix\url{http://arxiv.org/abs/hep-lat/9709102}.

\bibitem[{\citenamefont{Wennekers and Wittig}(2005)}]{Wittig:05:1}
\bibinfo{author}{\bibfnamefont{J.}~\bibnamefont{Wennekers}} \bibnamefont{and}
  \bibinfo{author}{\bibfnamefont{H.}~\bibnamefont{Wittig}},
  \bibinfo{journal}{J. High Energy Phys.} \textbf{\bibinfo{volume}{09}},
  \bibinfo{pages}{059} (\bibinfo{year}{2005}),
  \urlprefix\url{http://arxiv.org/abs/hep-lat/0507026}.

\bibitem[{\citenamefont{Shuryak and Verbaarschot}(1992)}]{Verbaarschot:92:1}
\bibinfo{author}{\bibfnamefont{E.~V.} \bibnamefont{Shuryak}} \bibnamefont{and}
  \bibinfo{author}{\bibfnamefont{J.~J.~M.} \bibnamefont{Verbaarschot}},
  \bibinfo{journal}{Nucl. Phys. A} \textbf{\bibinfo{volume}{560}},
  \bibinfo{pages}{306 } (\bibinfo{year}{1992}),
  \urlprefix\url{http://arxiv.org/abs/hep-th/9212088}.

\bibitem[{\citenamefont{Neuberger}(1998)}]{Neuberger:98:1}
\bibinfo{author}{\bibfnamefont{H.}~\bibnamefont{Neuberger}},
  \bibinfo{journal}{Phys. Lett. B} \textbf{\bibinfo{volume}{417}},
  \bibinfo{pages}{141} (\bibinfo{year}{1998}),
  \urlprefix\url{http://arxiv.org/abs/hep-lat/9707022}.

\bibitem[{\citenamefont{Bornyakov et~al.}(2007)\citenamefont{Bornyakov,
  Ilgenfritz, Martemyanov, Morozov, M\"{u}ller-Preussker, and
  Veselov}}]{Bornyakov:07:1}
\bibinfo{author}{\bibfnamefont{V.~G.} \bibnamefont{Bornyakov}},
  \bibinfo{author}{\bibfnamefont{E.}~\bibnamefont{Ilgenfritz}},
  \bibinfo{author}{\bibfnamefont{B.~V.} \bibnamefont{Martemyanov}},
  \bibinfo{author}{\bibfnamefont{S.~M.} \bibnamefont{Morozov}},
  \bibinfo{author}{\bibfnamefont{M.}~\bibnamefont{M\"{u}ller-Preussker}},
  \bibnamefont{and} \bibinfo{author}{\bibfnamefont{A.~I.}
  \bibnamefont{Veselov}}, \bibinfo{journal}{Phys. Rev. D}
  \textbf{\bibinfo{volume}{76}}, \bibinfo{pages}{054505}
  (\bibinfo{year}{2007}), \urlprefix\url{http://arxiv.org/abs/0706.4206}.

\bibitem[{\citenamefont{Kiskis and Narayanan}(2001)}]{Kiskis:01:1}
\bibinfo{author}{\bibfnamefont{J.}~\bibnamefont{Kiskis}} \bibnamefont{and}
  \bibinfo{author}{\bibfnamefont{R.}~\bibnamefont{Narayanan}},
  \bibinfo{journal}{Phys. Rev. D} \textbf{\bibinfo{volume}{64}},
  \bibinfo{pages}{117502} (\bibinfo{year}{2001}),
  \urlprefix\url{http://arxiv.org/abs/hep-lat/0106018}.

\bibitem[{\citenamefont{Verbaarschot}(1994)}]{Verbaarschot:94:2}
\bibinfo{author}{\bibfnamefont{J.~J.~M.} \bibnamefont{Verbaarschot}},
  \bibinfo{journal}{Nucl. Phys. B} \textbf{\bibinfo{volume}{426}},
  \bibinfo{pages}{559 } (\bibinfo{year}{1994}),
  \urlprefix\url{http://arxiv.org/abs/hep-th/9401092}.

\bibitem[{\citenamefont{Damgaard and Nishigaki}(2001)}]{Damgaard:01:1}
\bibinfo{author}{\bibfnamefont{P.~H.} \bibnamefont{Damgaard}} \bibnamefont{and}
  \bibinfo{author}{\bibfnamefont{S.~M.} \bibnamefont{Nishigaki}},
  \bibinfo{journal}{Phys. Rev. D} \textbf{\bibinfo{volume}{63}},
  \bibinfo{pages}{045012} (\bibinfo{year}{2001}),
  \urlprefix\url{http://arxiv.org/abs/hep-th/0006111}.

\bibitem[{\citenamefont{Gattringer}(2006)}]{Gattringer:06:1}
\bibinfo{author}{\bibfnamefont{C.}~\bibnamefont{Gattringer}},
  \bibinfo{journal}{Phys. Rev. Lett.} \textbf{\bibinfo{volume}{97}},
  \bibinfo{pages}{032003} (\bibinfo{year}{2006}),
  \urlprefix\url{http://arxiv.org/abs/hep-lat/0605018}.

\bibitem[{\citenamefont{Bilgici et~al.}(2008)\citenamefont{Bilgici, Bruckmann,
  Gattringer, and Hagen}}]{Gattringer:08:1}
\bibinfo{author}{\bibfnamefont{E.}~\bibnamefont{Bilgici}},
  \bibinfo{author}{\bibfnamefont{F.}~\bibnamefont{Bruckmann}},
  \bibinfo{author}{\bibfnamefont{C.}~\bibnamefont{Gattringer}},
  \bibnamefont{and} \bibinfo{author}{\bibfnamefont{C.}~\bibnamefont{Hagen}},
  \bibinfo{journal}{Phys. Rev. D} \textbf{\bibinfo{volume}{77}},
  \bibinfo{pages}{094007} (\bibinfo{year}{2008}),
  \urlprefix\url{http://arxiv.org/abs/0801.4051}.

\bibitem[{\citenamefont{{I. Horvath et al.}}(2003)}]{Horvath:03:1}
\bibinfo{author}{\bibnamefont{{I. Horvath et al.}}}, \bibinfo{journal}{Phys.
  Rev. D} \textbf{\bibinfo{volume}{68}}, \bibinfo{pages}{114505}
  (\bibinfo{year}{2003}), \urlprefix\url{http://arxiv.org/abs/hep-lat/0302009}.

\bibitem[{\citenamefont{{C. Aubin et al. [MILC
  Collaboration]}}(2005)}]{Aubin:04:1}
\bibinfo{author}{\bibnamefont{{C. Aubin et al. [MILC Collaboration]}}},
  \bibinfo{journal}{Nucl. Phys. Proc. Suppl.} \textbf{\bibinfo{volume}{140}},
  \bibinfo{pages}{626} (\bibinfo{year}{2005}),
  \urlprefix\url{http://arxiv.org/abs/hep-lat/0410024}.

\bibitem[{\citenamefont{Polikarpov et~al.}(2005)\citenamefont{Polikarpov,
  Gubarev, Morozov, and Zakharov}}]{Polikarpov:05:2}
\bibinfo{author}{\bibfnamefont{M.~I.} \bibnamefont{Polikarpov}},
  \bibinfo{author}{\bibfnamefont{F.~V.} \bibnamefont{Gubarev}},
  \bibinfo{author}{\bibfnamefont{S.~M.} \bibnamefont{Morozov}},
  \bibnamefont{and} \bibinfo{author}{\bibfnamefont{V.~I.}
  \bibnamefont{Zakharov}}, \bibinfo{journal}{PoS}
  \textbf{\bibinfo{volume}{LAT2005}}, \bibinfo{pages}{143}
  (\bibinfo{year}{2005}), \urlprefix\url{http://arxiv.org/abs/hep-lat/0510098}.

\bibitem[{\citenamefont{Koma et~al.}(2005)\citenamefont{Koma, Ilgenfritz,
  Koller, Schierholz, Streuer, and Weinberg}}]{Koma:05:1}
\bibinfo{author}{\bibfnamefont{Y.}~\bibnamefont{Koma}},
  \bibinfo{author}{\bibfnamefont{E.}~\bibnamefont{Ilgenfritz}},
  \bibinfo{author}{\bibfnamefont{K.}~\bibnamefont{Koller}},
  \bibinfo{author}{\bibfnamefont{G.}~\bibnamefont{Schierholz}},
  \bibinfo{author}{\bibfnamefont{T.}~\bibnamefont{Streuer}}, \bibnamefont{and}
  \bibinfo{author}{\bibfnamefont{V.}~\bibnamefont{Weinberg}},
  \bibinfo{journal}{PoS} \textbf{\bibinfo{volume}{LAT2005}},
  \bibinfo{pages}{300} (\bibinfo{year}{2005}),
  \urlprefix\url{http://arxiv.org/abs/hep-lat/0509164}.

\bibitem[{\citenamefont{Aharony et~al.}(2007)\citenamefont{Aharony,
  Sonnenschein, and Yankielowicz}}]{Aharony:07:1}
\bibinfo{author}{\bibfnamefont{O.}~\bibnamefont{Aharony}},
  \bibinfo{author}{\bibfnamefont{J.}~\bibnamefont{Sonnenschein}},
  \bibnamefont{and}
  \bibinfo{author}{\bibfnamefont{S.}~\bibnamefont{Yankielowicz}},
  \bibinfo{journal}{Ann. Phys.} \textbf{\bibinfo{volume}{322}},
  \bibinfo{pages}{1420} (\bibinfo{year}{2007}),
  \urlprefix\url{http://arxiv.org/abs/hep-th/0604161}.

\bibitem[{\citenamefont{Antonyan et~al.}()\citenamefont{Antonyan, Harvey,
  Jensen, and Kutasov}}]{Harvey:06:1}
\bibinfo{author}{\bibfnamefont{E.}~\bibnamefont{Antonyan}},
  \bibinfo{author}{\bibfnamefont{J.~A.} \bibnamefont{Harvey}},
  \bibinfo{author}{\bibfnamefont{S.}~\bibnamefont{Jensen}}, \bibnamefont{and}
  \bibinfo{author}{\bibfnamefont{D.}~\bibnamefont{Kutasov}},
  \emph{\bibinfo{title}{{NJL} and {QCD} from string theory}},
  \urlprefix\url{http://arxiv.org/abs/hep-th/0604017}.

\end{thebibliography}

\end{document}